# Effects of magnetic excitations and transitions on vacancy formation: cases of fcc Fe and Ni compared to bcc Fe

Kangming Li, Chu-Chun Fu

*Université Paris-Saclay, CEA, Service de Recherches de Métallurgie Physique, F-91191 Gif-sur-Yvette, France*

Anton Schneider

*Engineering Physics Department, University of Wisconsin, Madison WI 53706, USA*

Vacancy is one of the most frequent defects in metals. We study the impacts of magnetism on vacancy formation properties in fcc Ni, and in bcc and fcc Fe, via density functional theory (DFT) and effective interaction models combined with Monte Carlo simulations. Overall, the predicted vacancy formation energies and equilibrium vacancy concentrations are in good agreement with experimental data, available only at the high-temperature paramagnetic regime. Effects of magnetic transitions on vacancy formation energies are found to be more important in bcc Fe than in fcc Fe and Ni. The distinct behaviour is correlated to the relative roles of longitudinal and transversal spin excitations. At variance with the bcc-Fe case, we note a clear effect of longitudinal spin excitations on the magnetic free energy of vacancy formation in fcc Fe and Ni, leading to its steady variation above the respective magnetic transition temperature. Below the Néel point, such effect in fcc Fe is comparable but opposite to the one of the transversal excitations. Regarding fcc Ni, although neglecting the longitudinal spin excitations induces an overestimation of the Curie temperature by 220 K, no additional effect is visible below the Curie point. The distinct effects on the three systems are closely linked to DFT predictions of the dependence of vacancy formation energy on the variation of local magnetic-moment magnitudes and orientations.

## I. INTRODUCTION

Vacancy is one of the simplest and most frequently encountered structural defects in transition-metal systems. Vacancy concentration $[V]$ is a key parameter controling atomic transport via the vacancy mechanism[1]. The equilibrium monovacancy concentration $[V]_{eq}$ can be linked with the vacancy formation free energy $G_f^{\text{non-conf}}$ via the expression [2–4]

$$[V]_{eq} = exp(-\frac{G_f^{\text{non-conf}}}{k_B T}) \quad (1)$$

where $k_B$ and $T$ are respectively the Boltzmann constant and the absolute temperature, and $G_f^{\text{non-conf}}$ includes all the non-configurational entropy contributions (electronic, vibrational and magnetic) [5]. However, neither $[V]$ nor $G_f$ can always be directly measured in experiments. $[V]$ can be determined from differential dilatometry, but in a thermal-vacancy regime, the technique is only applicable at very high temperatures (sometimes near the melting point) due to the limited experimental resolution [6]. Other methods such as electrical resistivity measurement and positron annihilation spectroscopy usually provide the temperature dependence of $[V]$ rather than its absolute value [7]. For pure systems, $G_f^{\text{non-conf}}$ can also be indirectly estimated as $G_a - G_m$, where $G_a$ is the diffusion activation free energy determined from tracer diffusion experiments [8, 9], and $G_m$ is the vacancy migration free energy coming from resistivity recovery, internal friction or magnetic after-effect experiments [10–13].

In magnetic metal systems, several thermodynamic and kinetic properties are affected by thermal magnetic excitations and the magnetic transition. The additional magnetic degree of freedom makes an accurate prediction of those properties more difficult, particularly from an atomistic-simulation point of view. Experimental measurements of self- and solute diffusion coefficients in bcc Fe revealed strong effects of magnetism [9, 14–17]. These experimental evidences have motivated various recent theoretical studies [18–24]. However, very little is known about such effects on the vacancy-related properties in other magnetic systems (for example fcc Fe, fcc Ni and bcc Cr) that exhibit important longitudinal spin fluctuations at finite temperatures [25–28].

The thermal magnetic effects can be very system-dependent. Here we focus on fcc Fe and fcc Ni, which are the major constituents for the technologically important austenitic steels. At variance with the bcc-Fe case, experimental data on the vacancy concentration and the vacancy formation free energy are only available for the paramagnetic (PM) states of fcc Fe [29–33] and fcc Ni [6, 34–36], due to the low magnetic transition temperatures (67 K for fcc Fe [37] and 627 K for fcc Ni [38]). On the theoretical side, previous first principles investigations on the vacancy formation and diffusion properties in the PM regime were often informed by results on the ferromagnetic (FM) state of fcc Ni [39–43], and the nonmagnetic (NM) state in the case of fcc Fe [44–46]. It is still an open question whether these results, calculated with ordered or nonmagnetic states, are representative for those in the PM state. For example, the vacancy formation energy obtained for NM fcc Fe is 2.37 eV, much higher than the experimental values (1.40-1.83 eV [29–33]). More advanced descriptions of PM state properties can be achieved via the disordered local moment (DLM) approach [47, 48], the spin-wave method [49], spin dy-namics [50–52], spin-lattice dynamics [53, 54] and Monte Carlo simulations based on magnetic model Hamiltonians [55–58] (see Ref. [59] for a recent review), etc. Most of these studies [47–58, 60, 61] only considered defect-free systems. Regarding the vacancy properties in the PM state, most of the investigations addressed bcc Fe [20–24], while there are very few such studies of fcc Fe [62] or fcc Ni. Furthermore, the effects of longitudinal spin fluctuations, which can be important in the PM regime, are often neglected. Taking into account such effects requires, for example in the DLM approach, a more sophisticated statistical treatment combined with constrained local-moment calculations [25, 28, 63], leading to very CPU-demanding calculations. Finally, the previous theoretical studies in fcc Fe and Ni addressed the vacancy properties in the magnetic ground state and the high-temperature ideal PM state, whereas a systematic understanding of the temperature-dependent magnetic effects, which involve simultaneous longitudinal and transversal spin excitations, is missing.

In this study, we intend to provide a detailed analysis of the effects of longitudinal and transversal spin excitations and the magnetic transitions on vacancy formation properties, based on DFT parametrized magnetic effective interaction models (EIMs). The investigations are performed on fcc Fe and fcc Ni, using the EIMs parametrized in this work, and on bcc Fe with the EIM from a previous study [24]. A systematic comparison of the effects between the three systems is performed.

The paper is organized as follows. We introduce the computational details in Sec. II, and demonstrate the accuracy of the the EIMs in Sec. III. The vacancy formation properties from first principles calculations and the EIM predictions are discussed in Sec. IV, for various temperature regimes across the magnetic transition. Finally, conclusions are given in Sec. V.

## II. COMPUTATIONAL DETAILS

The present modelling strategy is as follows. We first performed DFT calculations in fcc Fe and Ni systems, with and without a vacancy (Sec. II A). Then, these results were used to parametrize EIMs (Sec. II B), which were later employed in on-lattice Monte Carlo simulations (Sec. II C), for the study of temperature-dependent vacancy formation properties.

### A. DFT calculations

DFT calculations were performed using the projector augmented wave (PAW) method [64, 65] as implemented in the Vienna Ab-initio Simulation Package (VASP) code [66–68]. $3d$ and $4s$ electrons of Fe and Ni atoms were considered as valence electrons. The plane-wave basis cutoff was set to 400 eV. The Methfessel-Paxton broadening scheme with a smearing width of 0.1 eV was used [69]. The convergence cut-off for the electronic self-consistency loop was set to $10^{-6}$ eV. The $k$-point grids were adjusted according to the cell size, to achieve a $k$-sampling equivalent to a cubic unit cell with a $16^3$ shifted grid following the Monkhorst-Pack scheme [70]. Atomic magnetic moments were obtained by an integration of spin-up and spin-down charge densities within the PAW spheres, with a radius of 1.302 Å for Fe and 1.286 Å for Ni.

We mainly adopted the generalized gradient approximation (GGA) for the exchange-correlation functional in the Perdew-Burke-Ernzerhof (PBE) parametrization [71]. Some additional local density approximation (LDA) calculations were performed for fcc Ni to examine the effects of the exchange-correlation functional on the vacancy formation energies.

We use the following notations for different magnetic states: NM, FM and PM denote nonmagnetic, ferromagnetic and paramagnetic states, respectively; AFS and AFD denote the antiferromagnetic single-layer and double-layer states, respectively; Z1 denotes the magnetic structure consisting of 3 consecutive spin-up layers and 1 spin-down layer along one direction. The PM state in DFT is approximated by magnetic special quasirandom structures (mSQS) [59].

DFT calculations were performed with various collinear magnetic structures for fcc Fe and Ni, and a few noncollinear ones for fcc Fe. For magnetically ordered states (NM, FM, AFS, AFD and Z1), the atomic positions, the volume, and the cell shape were fully optimized. For other magnetic states with local or global magnetic disorders, the atomic positions were fixed to those in the magnetic collinear ground states (namely AFD Fe and FM Ni) while the volume and the cell shape were optimized.

For the defect-free systems, supercells of various sizes containing up to 128 atoms were used. In total, we considered more than 120 and 60 different magnetic configurations for fcc Fe and Ni, respectively.

For the systems containing a vacancy, we performed calculations with either fully relaxing the magnetic configurations or constraining atomic magnetic moments. In the former case, 108-site and 128-site supercells were used. They were respectively constructed as 3×3×3 4-atom face-centred cubic (fcc) unit cells and 4×4×4 2-atom body-centred tetragonal (bct) unit cells. In the constrained-magnetism case, we used 54-site (3×3×3 2-atom unit cells) and 72-site (3×3×4 2-atom unit cells) supercells. The vacancy formation energy $E_f$ was calculated as

$$E_f = E_{tot,V} - \frac{N-1}{N} E_{tot,0} \quad (2)$$

with $E_{tot,V}$ and $E_{tot,0}$ the total energies of the systems with and without a vacancy. We verified that $E_f$ calculated for a given magnetic configuration but with different supercell sizes differs by at most 5 meV. In total, we considered around 140 and 90 different magnetic configurations (containing a vacancy) for fcc Fe and Ni, respectively.

Vibrational entropies of vacancy formation of the magnetic collinear ground states were calculated within the harmonic approximation in $3 \times 3 \times 3$ supercells using VASP and PHONOPY [72]. They were then added to the vacancy formation magnetic free energies from Monte Carlo simulations. With this approximation, the magnon-phonon coupling effects [21, 22, 73] were not considered.

### B. Effective interaction models (EIM)

We adopt a model Hamiltonian similar to those used to investigate magnetic properties, phase stability [55–57, 61] and vacancy formation and diffusion properties [24, 58?] of Fe-based systems. The Hamiltonians for the present pure systems are as follows:

$$E_{tot} = \sum_i \sigma_i \cdot (\epsilon_i + A_i M_i^2 + B_i M_i^4 + \sum_j \sigma_j \cdot J_{ij} \boldsymbol{M}_i \boldsymbol{M}_j) \quad (3)$$

where $i$ denotes the $i$-th lattice site, $\sigma_i$ is the occupation variable and is equal to 1 (0) for an occupied (empty) lattice site, $\epsilon_i$ is the on-site nonmagnetic parameter, $M_i$ is the local magnetic moment, $A_i$ and $B_i$ are the on-site magnetic parameters, $J_{ij}$ are the exchange interaction parameter and $\sum_j$ is a sum over all the neighbouring sites up to the fourth-neighbour shell. The Hamiltonian has a generalized Heisenberg form enabling both longitudinal and transversal spin variations. We note that the effects of the presence of a vacancy are incorporated as the dependence of the model parameters on the distance between the $i$-th atom and the vacancy [?].

In this work, we investigate the vacancy formation properties in fcc Fe and Ni, and compare them with those in bcc Fe. For bcc Fe, our previously developed model is used [24]. For fcc Fe and Ni, we parametrize new EIMs as follows. First, the magnetic parameters of bulk atoms are fitted to energy differences between various magnetic states of defect-free fcc Fe and Ni systems. Then, the magnetic parameters of the atoms in the first and second coordination shells of the vacancy are fitted to the energy differences between different magnetic structures containing a vacancy. Finally, the nonmagnetic parameters are fitted to the vacancy formation energies of AFD Fe and FM Ni (the respective collinear magnetic ground states). The numerical values of the parameters can be found in the Supplemental Material [?].

### C. On-lattice Monte Carlo simulations

On-lattice spin Monte Carlo (SMC) simulations for fcc Fe and Ni are performed on fcc supercells consisting of $16^3$ cubic unit cells (16384 fcc sites) with periodic boundary conditions. For bcc Fe, we use bcc supercells consisting of $20^3$ cubic unit cells (16000 bcc sites).

The spin system is equilibrated at a given temperature by performing Metropolis SMC steps. At each SMC step, a random variation of the magnetic moment of a randomly chosen atom is attempted with the acceptance probability $min[1, exp(\Delta E/\eta)]$. Within the classical statistics, $\eta$ is equal to $k_B T$. However, a quantum statistics treatment of spins at low temperatures was shown to be necessary for various magnetic and vacancy properties [24]. For the quantum treatment of spins, we adopt the methodology described in Ref. [74], where the scaling factor $\eta$ below the magnetic transition temperature is determined based on the temperature-dependent magnon density of states (mDOS). In this work, the mDOS is calculated within the quasiharmonic approximation (QHA). Please note that the QHA is expected to be accurate only at low temperatures, and it could lead to a sharp transition from quantum to classical statistics if applied to higher temperatures [54, 74]. In practice, we observe that this quantum-classical transition is in particular sharp in fcc Ni because of its very high magnon energy.

In principle, the vacancy formation energy $E_f$ can be calculated in a brute-force manner by Eq. 2, using the thermodynamic averages of the total energies for the systems with and without a vacancy. However, it is extremely computationally inefficient to achieve an desired accuracy (e.g., 1 meV) due to the large fluctuations in the total energies. Furthermore, at finite temperatures, we need to determine the magnetic entropy and free energy of vacancy formation, respectively $S_f$ and $G_f$. Their calculation from the integration of $E_f$ requires a even higher accuracy of the $E_f$. In this work, we use the efficient Monte Carlo schemes detailed in the Supplemental Material [75?] to calculate directly $G_f$ with a satisfactory accuracy (<1 meV). Once $G_f$ is obtained, $S_f$ and $E_f$ can be easily evaluated from usual thermodynamic relations. We verify that these schemes yield the same results as those from the aforementioned brute-force method [?], and one of them was already successfully applied to the study of bcc Fe [24].

### III. ACCURACY OF THE EIMs

As shown in Table I, the energetic hierarchy of several collinear magnetic states of fcc Fe from the EIM is in good agreement with the DFT results. The magnetic ground state of Fe predicted by the EIM is a spin spiral with $\boldsymbol{q} = \frac{2\pi}{a}(0, 0, 0.3)$, which has a slightly lower energy than the AFD state.

Experimentally, the magnetic ground state of fcc Fe is a spin spiral with $\boldsymbol{q} = \frac{2\pi}{a}(0.3, 0, 1)$ [76–78]. First principles calculations also predicted a spin spiral ground state, but with $\boldsymbol{q} = \frac{2\pi}{a}(0, 0, 0.6)$ [79–81]. Within the collinear approximation, the ground state is the AFD state [81–

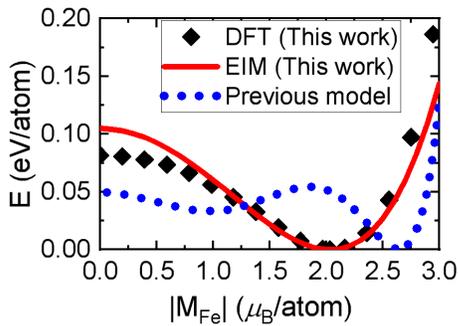

FIG. 1. (Color online) Energy of AFD fcc Fe as a function of spin magnitude from the DFT and EIM predictions in this study and the prediction from a previous model [55].

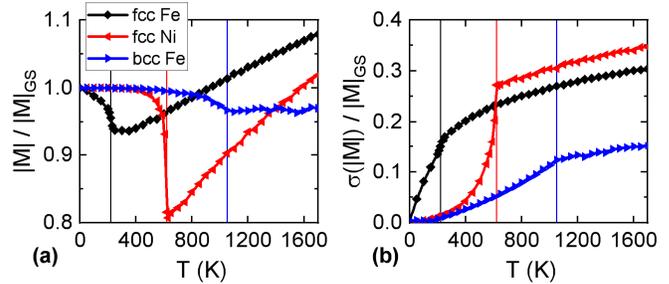

FIG. 2. (Color online) Variation with temperature of (a) average spin magnitudes and (b) standard deviation of the spin magnitudes, scaled by the respective ground-state moments $|M|_{GS}$. The vertical lines denotes the corresponding magnetic transition temperatures.

84], with $\boldsymbol{q} = \frac{2\pi}{a}(0, 0, 0.5)$.

We fit the EIM parameters of fcc Fe on the DFT results of various magnetic states, in particular on the energy of the AFD state as a function of magnetic-moment magnitude (Fig. 1). The latter energy landscape is not well captured by a recent effective interaction model in Ref. [55], which also gives an overestimated equilibrium spin magnitude for the AFD Fe. Note that it is crucial to well reproduce such an energetic landscape, because it dictates the description of longitudinal spin excitations at finite temperatures, which are important in fcc Fe and Ni [26–28]. Furthermore, our DFT results show that the vacancy formation energies are sensitive to the spin magnitudes.

The EIM of pure Ni is also fitted on the DFT results of various collinear states and the energy-versus-moment relation in the FM state. In agreement with the DFT results, the magnetic ground state of Ni predicted by the EIM is FM with a magnetic moment of 0.65 $\mu_B$.

TABLE I. Energies (in meV/atom) of several magnetic states (with respect to AFD) of fcc Fe by the DFT and EIM predictions in this study and the predictions from a previous model [55].

|  | Noncol-AF | AFD | AFS | FM | NM |
|---|---|---|---|---|---|
| DFT | - | 0 | 19 | 22 | 82 |
| EIM | -0.5 | 0 | 17 | 29 | 105 |
| EIM in Ref. [55] | -2 | 0 | 0 | 32 | 51 |

## IV. RESULTS AND DISCUSSIONS

In this section, we present and discuss the DFT and EIM predictions of bulk and vacancy formation properties in fcc Fe and Ni, with a comparison to those in bcc Fe.

### A. Magnetic properties of the defect-free bulk systems

As shown in Table II, the generalized Heisenberg (GH) EIMs reproduce well the experimental Curie temperatures of bcc Fe and fcc Ni, and give a closer prediction of the measured Néel temperature of fcc Fe than that of 450 K from a previous model [55]. Though the predicted Néel temperature is still somehow overestimated, it is ensured that fcc Fe is already paramagnetic at the room temperature.

TABLE II. Experimental and calculated Curie temperature of bcc Fe and fcc Ni and Néel temperature of fcc Fe. GH and CH denote the generalized Heisenberg (i.e., freely evolving spin magnitudes) and classical Heisenberg (i.e., fixed spin magnitudes) simulations, respectively

| System | Exp. | GH | CH |
|---|---|---|---|
| bcc Fe | 1044 K [38] | 1050 K | 1080 K |
| fcc Fe | 67 K [37] | 220 K | 240 K |
| fcc Ni | 627 K [38] | 620 K | 880 K |

It is known that longitudinal spin fluctuations are more significant in fcc Fe and Ni than in bcc Fe [26–28]. This point is also confirmed by the EIM results shown in Fig. 2. Compared to the bcc-Fe case, fcc Ni and fcc Fe show a stronger variation of the average spin magnitude versus temperature, and a larger dispersion of the spin magnitudes.

Our generalized Heisenberg EIMs naturally account for longitudinal spin fluctuations. To demonstrate the effects of such fluctuations on the magnetic transition temperatures $T_C$, we also perform classical Heisenberg (CH) simulations based on the same EIMs, by constraining the spin magnitudes to the ground-state values. The difference between $T_C^{GH}$ and $T_C^{CH}$ is then due to the fact that the temperature-dependent variation of the average spin magnitude, and the dispersion of spin magnitudes are neglected in CH simulations. The effects of neglecting the variation in the average spin magnitude can be roughly estimated as follows. Since the magnetic transition temperature is proportional to $J_{ij}M^2$, $T_C^{CH}$ can be estimated as

$$T_C^{CH} = T_C^{GH} \times \frac{M^2(0K)}{M^2(T_C^{GH})} \quad (4)$$

where $M^2(0K)/M^2(T_C^{GH})$ according to Fig. 2(a) is equal to 1.07, 1.14 and 1.53 for bcc Fe, fcc Fe and fcc Ni, respectively. These estimations are consistent with the results obtained from CH simulations (Table II). The actual difference in the transition temperatures between the GH and CH simulations is small for bcc and fcc Fe, but it becomes quite large in fcc Ni, showing the necessity of taking into account the temperature variation of spin magnitudes. On the other hand, if we impose the Ni spin magnitudes (in CH simulations) to be the average value from the GH simulations at the $T_C^{GH}$, the obtained $T_C^{CH}$ is 60 K lower than the $T_C^{GH}$. This suggests a minor but non-negligible effect of the dispersion of atomic spin magnitudes at a given temperature.

### B. Vacancy formation energies at 0 K from DFT

The vacancy formation energies $E_f$ in various magnetic states are computed from DFT (Table III). The dispersion in $E_f$ among different magnetic states is much larger in bcc Fe than in fcc Fe and Ni, suggesting a stronger magnetic effects on $E_f$ in bcc Fe. Please note that such dispersions are related to not only the ordering of local magnetic-moment orientations but also the atomic spin magnitudes. The latter can differ by as much as 0.9 $\mu_B$ between different magnetic states (excluding the NM one) in bcc and fcc Fe.

TABLE III. $E_f$ (in eV) in different magnetic states computed from DFT. The collinear magnetic ground state values are marked in bold.

| System | NM   | FM       | AFS  | AFD      | mSQS |
|--------|------|----------|------|----------|------|
| fcc Fe | 2.37 | 1.85     | 2.00 | **1.83** | 2.04 |
| fcc Ni | 1.38 | **1.43** | 1.47 | 1.53     | 1.52 |
| bcc Fe | 0.57 | **2.20** | 0.94 | 1.86     | 1.59 |

To check the dependence of $E_f$ on the atomic spin magnitudes $|M|$, we perform the DFT calculations by constraining all the local $|M|$ to the same values. The results for the three systems are shown in Fig. 3.

$E_f$ is less dependent on $|M|$ in fcc Ni than in fcc and bcc Fe. In fcc Ni, $E_f$ of the FM state is lower than most of other magnetic states with the same $|M|$. Consequently, $E_f$ in fcc Ni at finite temperatures is expected to be larger than that of the FM ground state.

In fcc Fe, $E_f$ for all the spin-orientation orderings tends to decrease with increasing $|M|$ for $|M|$ below $2\mu_B$. Fig. 3 also shows that the difference of $E_f$ is relatively small between various spin orderings with the same $|M|$, whereas the variation in $E_f$ with $|M|$ is relatively large for a given spin ordering. This suggests $E_f$ in fcc Fe has a stronger dependence on the atomic spin magnitudes than on the spin orientations. In particular, consider the AFD state and the mSQS. According to the results obtained with optimized magnetic moments (Table III), $E_f$ in the AFD state is 0.2 eV lower than that in the mSQS. From Fig. 3, it is clear that $E_f$ of the mSQS and the AFD state is very similar for the same $|M|$. Therefore, the difference of $E_f$ between the two states in Table III is mainly due to the fact that the optimized $|M|$ is smaller in the mSQS (on average 1.5 $\mu_B$) than in the AFD state (2.0 $\mu_B$). This indicates that $E_f$ in the mSQS with optimized magnetic moments cannot be taken as the value in the PM state, since longitudinal spin fluctuations in the high temperature PM state lead to an increase in $|M|$.

Contrary to the fcc-Fe case, $E_f$ in bcc Fe is much more dependent on the spin-orientation orderings than on the atomic spin magnitudes, as demonstrated in Fig. 3(c). Furthermore, the variation in $|M|$ is also expected to be small in bcc Fe: the optimized magnetic moments in the mSQS are on average 2.05 $\mu_B$, rather close to 2.20 $\mu_B$ in the FM ground state; it is also shown in the previous section that the thermal longitudinal spin fluctuations in bcc Fe are small. Therefore, the temperature-dependent vacancy formation properties in bcc Fe are expected to be mainly determined by the arrangement of spin orientations.

We note that our EIMs reproduce satisfactorily the above DFT predictions in Table III and Fig. 3.

Finally, we would like to comment on the effects of exchange-correlation functionals on $E_f$. In this work, we parameterize the EIMs of fcc Fe and Ni based on the GGA-PBE results. It is well known that GGA describes the bulk (in particular magnetic) properties of Fe better than LDA [81]. For Ni, LDA and GGA equilibrium lattice parameters ($a_0$) are 3.416 Å and 3.514 Å, respectively, the latter being closer to the experimental value of 3.52Å [85]. By performing LDA calculations using the GGA $a_0$, we verified that adopting LDA results only introduced a constant shift in $E_f$ for all the magnetic states compared with GGA values, namely the relative difference in $E_f$ between different magnetic states is the same using LDA and GGA. Therefore, the choice of exchange-correlation functionals influences only the parametrization of the nonmagnetic part of the EIMs. This point will be further discussed for Ni in Sec. IV C 1.

### C. Vacancy formation properties at finite temperatures from EIM

We show in Fig. 4 the temperature-dependent vacancy formation properties calculated from GH simulations for the three systems. Here, $E_f^{mag}$ is the vacancy formation energy, $S_f^{mag}$ is the magnetic contribution to the vacancy formation entropy, and $G_f^{mag}$ ($=E_f^{mag} - TS_f^{mag}$) is the vacancy formation magnetic free energy (without the vibrational or other entropic contributions). In the following, we first describe the overall behaviour versus temperature and compare our results with experimental data. Then, the effects of magnetism in different temperature ranges are analysed separately.

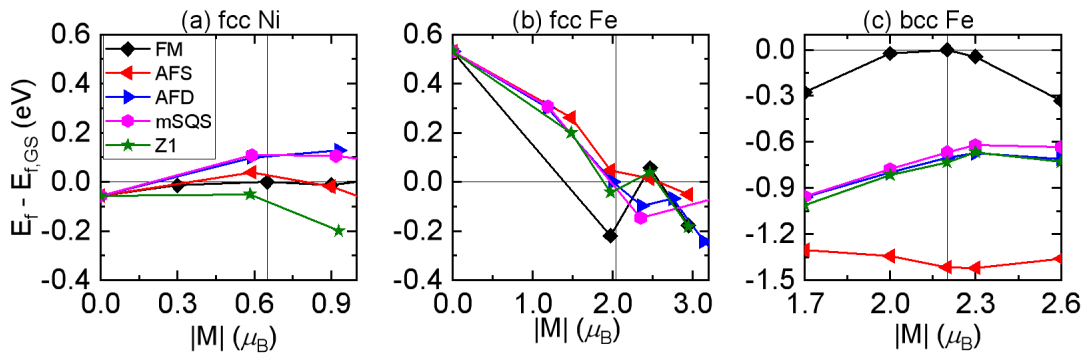

FIG. 3. (Color online) DFT results of $E_f$ in the three systems as a function of constrained local magnetic moment, with respect to $E_f$ in the collinear magnetic ground states (1.43 eV for FM fcc Ni, 1.83 eV for AFD fcc Fe and 2.20 eV for FM bcc Fe). The vertical lines denote the spin magnitude of the magnetic ground states.

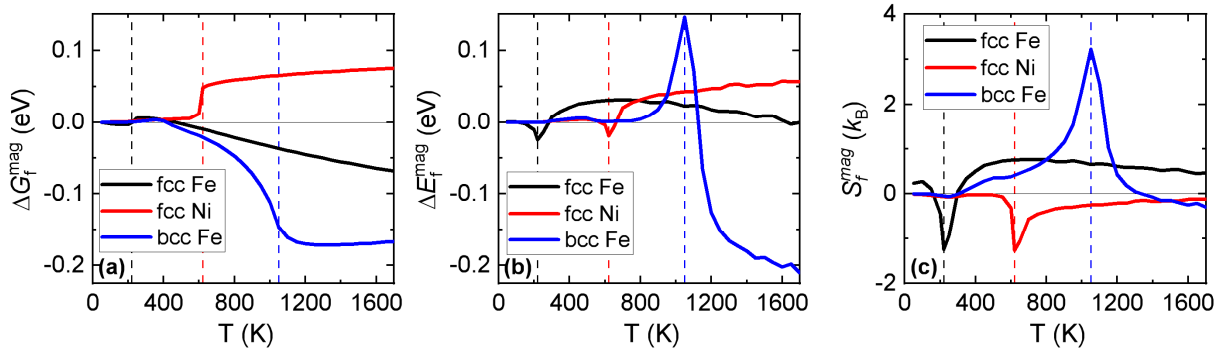

FIG. 4. (Color online) Magnetic contribution to vacancy formation properties in the three systems calculated from GH simulations (i.e., with both transversal and longitudinal spin fluctuations). $G_f^{mag}$ and $E_f^{mag}$ are given with respect to the ground-state values (1.43 eV for FM fcc Ni, 1.83 eV for AFD fcc Fe and 2.20 eV for FM bcc Fe). The vertical lines denotes the EIM-predicted magnetic transition temperatures.

### 1. Overall behaviour and comparison with experiments

We first consider bcc Fe in which the vacancy activation energy strongly depends on the magnetic state as evidenced by experiments [9, 86]. Theoretical studies reveal that both vacancy formation and migration energies are lower in the fully PM than in the FM state [18–24]. As can be seen in Fig. 4, compared with the FM values, the asymptotic $G_f^{mag}$ and $E_f^{mag}$ are reduced by 0.17 and 0.20 eV, respectively.

At variance with the well-known case of bcc Fe, magnetic disorders in fcc Fe and Ni do not lead to a strong decrease of $G_f^{mag}$. In the PM region close to the magnetic transition, $\Delta G_f^{mag} = G_f^{mag}(T) - G_f^{mag}(0K)$ is negligible in fcc Fe but it is positive in fcc Ni. Furthermore, there is a steady variation in $G_f^{mag}$ of fcc Fe and Ni in the PM region, which is different from the saturation behaviour observed in bcc Fe. These distinct behaviours will be further analysed in Sec. IV C 3.

The vacancy formation energies in the magnetic ground state and the PM state are compared with other calculations and experiments in Table IV. In bcc Fe, our FM value agrees with the previous calculated results ranging between 2.15-2.23 eV based on the same exchange-correlation functional GGA-PBE [18, 22, 23], whereas the GGA-PW91 prediction gives a lower value of 1.95 eV [82], or 2.13 eV if using the experimental $a_0$ [19]. Our fully PM value is consistent with the previous results, and the dispersion in the calculated values may be due to the different atomic-position relaxation schemes [19–23]. Before comparing with experimental data, it should be noted that the measurement of $E_f$ in bcc Fe is extremely sensitive to the presence of interstitial impurities such as carbon, which could lead to an underestimation of $E_f$ in earlier studies [86]. Compared with the experimental data in bcc Fe, both our ground-state and PM values are on the higher limit. A very relevant quantity to compare is the difference between the GS and PM energies. It summarizes the overall effect of the mag-netic transition, and a cancellation of systematic errors can occur in both calculations and experiments. This difference from our prediction is in good agreement with the result of Schepper et al. [86].

TABLE IV. Vacancy formation energies (in eV) in the collinear magnetic ground state and the PM state from calculations and experiments. For the PM state, we show both $E_f^{mag}$ and $G_f^{mag}$ (given inside the parentheses) calculated at 1500 K. The small difference between $E_f^{mag}$ and $G_f^{mag}$ comes from the longitudinal spin entropy as explained in Sec. IV C 3.

|  | bcc Fe | | fcc Fe | | fcc Ni | |
| --- | --- | --- | --- | --- | --- | --- |
|  | FM | PM | AFD | PM | FM | PM |
| This work | 2.20 | 2.00 (2.03) | 1.84 | 1.85 (1.78) | 1.43 | 1.48 (1.50) |
| Other calculations | 1.95-2.24 [18–23, 82] | 1.54-1.98 [19–23] | 1.82 [83] | 1.86 [62] | 1.43 [39] |  |
| Experiments | 2.0±0.2 [86] | 1.79±0.1 [86] |  | 1.40±0.15 [33] |  | 1.73±0.07 [36] |
|  | 1.81±0.1 [87] | 1.74±0.1 [87] |  | 1.7±0.2 [32] |  | 1.7±0.1 [35] |
|  | 1.60±0.15 [29] | 1.53±0.15 [29] |  | 1.83±0.14 [30] |  | 1.6±0.1 [35] |
|  |  |  |  |  |  | 1.56±0.04 [?] |

For fcc Fe, our results agree with the previous ab initio predictions [62, 83]. On the experimental side, the measurements on $E_f$ are performed in the PM state. Unlike the bcc-Fe case, the measurement in fcc Fe is less affected by the presence of impurities [29]. The experimental uncertainty is rather due to its limited temperature window of stability [33]. The predicted $E_f$ of PM fcc Fe is consistent with the measured values within the experimental uncertainty. We note that this value is much lower than the one in NM fcc Fe (2.37 eV). The NM state is therefore a poor representative phase to study diffusion properties in fcc Fe, although it has been used in some recent ab initio studies on diffusions in PM fcc Fe [44–46].

To the best of our knowledge, there is no theoretical result for the PM $E_f$ and no experimental data for the FM $E_f$ in fcc Ni. The experimental $E_f$ ranges between 1.5 eV and 1.8 eV (see Ref. [5] and references therein). Smedskjaer et al. [34] suggested that the experimental discrepancies arise from the different analysis methods and the associated assumptions between positron annihilation spectroscopy experiments, and uncontrolled metallurgical variables for other techniques such as positron lifetime spectroscopy. In the light of this suggestion, the recommended value of $E_f$ in Ref [6] is 1.79 eV. The results from the recent experiments [35?, 36] are shown in Table IV. Though being higher than the recommended value in Ref [6], our prediction of $E_f$ for PM Ni is within the uncertainty of the experiments [33, 35, 88?], including the most recent one using differential dilatometry [?] to measure directly the equilibrium vacancy concentration.

Indeed, the experimental $E_f$ is indirectly deduced from the temperature dependence of the equilibrium vacancy concentrations. Therefore, the latter may allow a more direct comparison. To the best of our knowledge, such measurements on vacancy concentrations have been performed in fcc Ni, but not in bcc and fcc Fe. To determine these concentrations, we have calculated via DFT the vacancy vibrational formation entropy for FM fcc Ni (2.15 $k_B$), and included it to the magnetic free energy of vacancy formation obtained using the EIM. As shown in Fig. 5, the calculated equilibrium vacancy concentrations of fcc Ni agree well with the experimental concentrations, which exhibit a larger dispersion at temperatures below 1400 K. We note that $E_f$ deduced as the slope of the experimental curve is very sensitive to this dispersion of the experimental data: the estimated $E_f$ ranges between 1.5 to 1.6 eV if only data above 1400 K are considered, and between 1.45 to 1.80 eV if all data are considered.

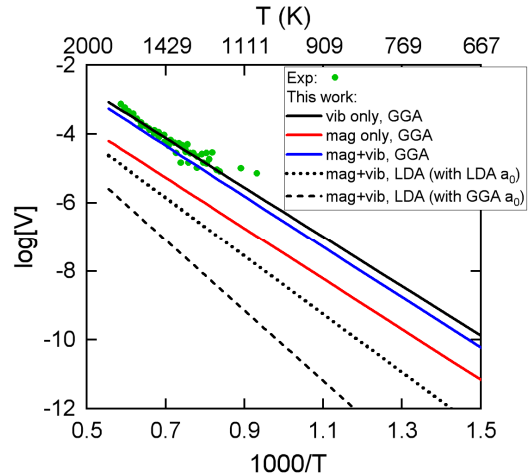

FIG. 5. (Color online) Calculated $[V]_{eq}$ as a function of temperature, compared to available experimental data in fcc Ni [89?–91].

On the theoretical side, $E_f$ of FM Ni is sensitive to the choice of exchange-correlation functional, for instance between the GGA-PBE and the LDA functionals used in DFT calculations [92]. As discussed in Sec. IV B, the choice affects only the EIM prediction of $E_f$ of FM Ni, but not the temperatures evolution $\Delta G_f^{mag}(T)$. To evaluate the LDA-based prediction of equilibrium vacancy concentrations, we calculated the vacancy formation energy (1.65 eV) and the vacancy vibrational formation entropy (0.43 $k_B$) of FM Ni via DFT using LDA, and combine them with $\Delta G_f^{mag}(T)$ predicted by the EIM. Since the experimental $a_0$ of Ni is well reproduced by GGA-

PBE but underestimated by LDA, we also applied LDA with the GGA-PBE $a_0$ to calculate the vacancy formation energy (2.02 eV) and the vacancy vibrational formation entropy (0.44 $k_B$) for FM Ni. We note the latter is much lower than the GGA-based value (2.15 $k_B$) and the experimental result (3.3 ±0.5 $k_B$ [?]). As shown in Fig. 5, the two LDA-based predictions substantially underestimate the experimental equilibrium vacancy concentrations. This may be due to the large vacancy formation energy and the the small vacancy vibrational formation entropy obtained with the LDA functional. Therefore, we conclude that the GGA-based prediction is more consistent with the experimental vacancy concentrations.

Finally, we also compare the relative importance of the vibrational and magnetic effects on the equilibrium vacancy concentration in fcc Ni. As can be seen from Fig. 5, neglecting the magnetic effects (namely $G_f = G_f^{mag}(0K) - TS_f^{vib}$) changes very little the predicted vacancy concentrations, whereas neglecting the vibrational contribution (namely $G_f = G_f^{mag}(T)$) would underestimate the vacancy concentration by one order of magnitude. In fact, for the three systems, the vibrational contribution to the vacancy concentration determination is stronger than the magnetic contribution. The latter is small in fcc Fe and Ni, but it is more significant in bcc Fe as neglecting this can lead to an underestimation of vacancy concentration by up to a factor of 5.

#### 2. Vacancy formation below magnetic transitions

The EIMs for bcc Fe and fcc Ni predict the same FM ground state as with DFT, therefore the ground-state $E_f$ for bcc Fe and fcc Ni are correctly reproduced. For fcc Fe, the collinear magnetic ground state by DFT is AFD, whereas the EIM predicts a spin spiral ground state with a slightly lower energy (by 0.5 meV/atom) than the AFD state. The $E_f$ of the spin spiral is found to be slightly higher (by 0.01 eV) than the AFD value.

The vacancy formation properties below the magnetic transitions can be influenced by the transversal and the longitudinal spin fluctuations, and the effects of two types of fluctuations cannot be easily decoupled. In Fig. 6, we compare the temperature evolution of $\Delta G_f^{mag}$ predicted by GH and CH simulations.

Below the magnetic transitions, the CH results (without longitudinal spin fluctuations) follow the same trend as the GH values for bcc Fe and fcc Ni, indicating the effects of the longitudinal spin fluctuations are small. This is expected for bcc Fe in which longitudinal spin fluctuations are not significant. For fcc Ni, longitudinal spin fluctuations are rather strong, and the Curie temperature (880 K) from the CH simulations is much higher than the GH value (620 K). However, the CH and GH results of $\Delta G_f^{mag}$ would be similar if they are rescaled to the same Curie temperatures. This is in agreement with the DFT results in Sec. IV B where $E_f$ in fcc Ni is found to be rather insensitive to the spin magnitudes, especially for the FM state.

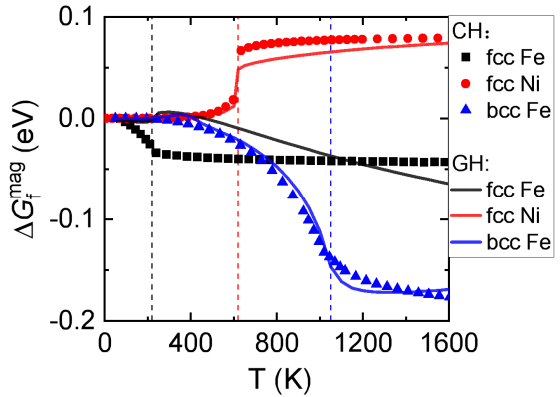

FIG. 6. (Color online) Comparison of $\Delta G_f^{mag}(= G_f^{mag}(T) - G_f^{mag}(0K))$ as a function of temperature between the GH and CH values. For each pure system, the CH curve is rescaled to have the same magnetic transition temperature as the GH curve. The vertical lines denote the magnetic transition temperatures of the GH curves. $T_{\text{Néel}}$ of fcc Fe is 220 K (240 K) from the GH (CH) calculations; $T_{\text{Curie}}$ of fcc Ni is 620 K (880 K) from the GH (CH) calculations; $T_{\text{Curie}}$ of bcc Fe is 1050 K (1080 K) from the GH (CH) calculations.

For fcc Fe below the Néel temperature, the CH values decrease with temperature while the GH values remain nearly unchanged. This suggests that below the magnetic transition in fcc Fe, the transversal spin fluctuations lead to a decrease in $G_f^{mag}$, while the longitudinal spin fluctuations lead to an increase in $G_f^{mag}$. Indeed, the spin magnitudes tend to decrease with temperature before the Néel temperature (Fig. 2), which leads to an increase in $G_f^{mag}$ according to our DFT results in Fig. 3(b).

On the other hand, the spin magnitudes tend to increase with temperature above the Néel temperature, so $G_f^{mag}$ is expected to decrease with temperature according to the DFT results in Fig. 3(b). As shown later, there is a steady decrease in $G_f^{mag}$ with temperature in PM fcc Fe. This demonstrates that the longitudinal spin fluctuations play a dominant role over the transversal ones in the temperature evolution of $G_f^{mag}$ of fcc Fe.

#### 3. Vacancy formation properties in the PM regime

As shown in Fig. 6, $G_f^{mag}$ always saturates in the PM region in the CH simulations. From thermodynamic relations, it can thus be concluded that $E_f^{mag}$ also saturates while $S_f^{mag}$ vanishes at high $T$. Indeed, for CH models, the magnetic contribution to the vacancy formation energy is zero in the ideal PM state, since the thermodynamic average of the spin-spin correlation $<\bm{M}_i\bm{M}_j>$ is zero in the ideal PM state.

The steady variation in $G_f^{mag}$ in the PM region in Fig. 4(a) is thus related to the longitudinal spin fluctuations and is observed only with the GH models. This behaviour can be quantitatively understood as follows. In the ideal PM region where the spin-spin correlations are negligible, the magnetic Hamiltonian can be approximated by a simplified form, retaining only the on-site terms, namely

$$E_{tot}^{mag} = \sum_i e_i^{mag} = \sum_i A_i M_i^2 + B_i M_i^4 \quad (5)$$

where $e_i^{mag} = A_i M_i^2 + B_i M_i^4$ is the on-site magnetic energy for the $i$-th atom. The partition function can be expressed as

$$Z_{tot} = \int e^{-\beta E_{tot}^{mag}} d\boldsymbol{M}_1 ... d\boldsymbol{M}_N = \prod_i z_i \quad (6)$$

$$z_i = \int e^{-\beta e_i^{mag}} d\boldsymbol{M}_i \quad (7)$$

The total magnetic energy of the system can then be calculated as

$$G_{tot}^{mag} = -k_B T \cdot ln Z_{tot} = \sum_i g_i^{mag} \quad (8)$$

$$g_i^{mag} = -k_B T ln z_i \quad (9)$$

where $g_i^{mag}$ is the magnetic free energy of the $i$-th atom. In our EIMs, there are 3 sets of the on-site parameters for each pure system depending on the distance from the vacancy. Consequently, there are three possible values of $g_i^{mag}$ for a pure system containing a vacancy. We use the subscript $i$ equal to 0, 1, and 2 to represent the bulk atoms and those in the first and second neighbouring shells of the vacancy, respectively. The vacancy formation free energy can then expressed as

$$G_f^{mag} = G_{tot,V}^{mag} - \frac{N-1}{N} G_{tot,0}^{mag}$$
$$= \sum_{i=1,2} n_i \cdot (g_i^{mag} - g_0^{mag}) \quad (10)$$

where $n_i$ is the coordination number of the $i$-th shell.

The simplified EIMs allow a direct numerical evaluation of $z_i$ and thus all the subsequent quantities. It should be noted that even though the contribution of transversal spin fluctuations to the averaged total energy is negligible in the PM state, its contribution to the total entropy is not zero. This is fully taken into account in Eq. 7, where it is equally possible for $\boldsymbol{M}_i$ to take any direction, as expected for the perfectly PM state. In this way, the transversal part of the entropy per atom is maximized and the same for all the atoms independent of their distances from the vacancy. This entropic contribution is thus cancelled out in Eq. 10.

The results of $G_f^{mag}$ and $S_f^{mag}$ are compared between the complete and the simplified EIMs for fcc Fe and Ni in Fig. 7. In the PM region, $G_f^{mag}$ and $S_f^{mag}$ using the simplified EIMs converge to those from the complete EIMs, confirming that the simplified EIMs are very good representations of the complete ones at high temperatures.

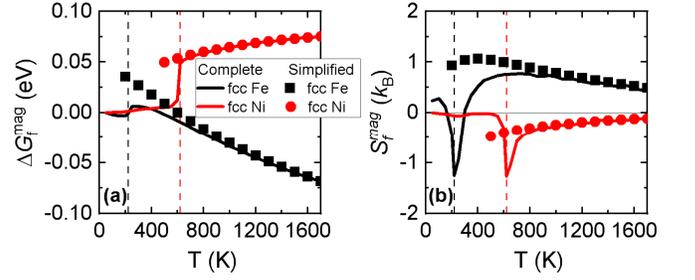

FIG. 7. (Color online) (a) $\Delta G_f^{mag}$ and (b) $S_f^{mag}$ predicted from the complete and simplified EIMs. The vertical lines denote the EIM-predicted magnetic transition temperatures.

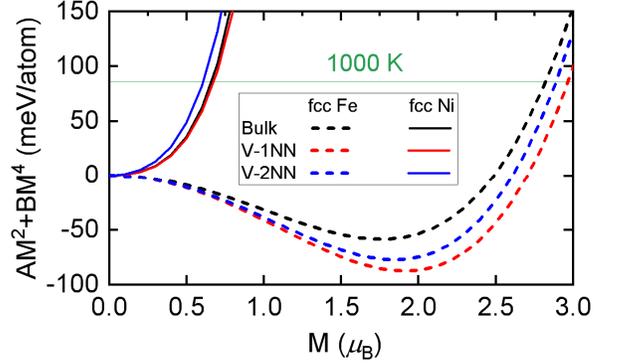

FIG. 8. (Color online) On-site magnetic energies $A_i M_i^2 + B_i M_i^4$ for the bulk atoms and first and second nearest neighbours of a vacancy.

The different variation trends in $G_f^{mag}$ between fcc Fe and Ni are related to the signs of $S_f^{mag}$. This can also be understood with the simplified Hamiltonians that contain only the on-site magnetic energy $A_i M_i^2 + B_i M_i^4$. As shown in Fig. 8, for a given temperature (e.g. 1000 K as indicated in the figure), the atoms around the vacancy can explore a larger interval of $|M|$, and thus has a larger entropy than the bulk atoms in fcc Fe, while the case is reversed for fcc Ni.

The current analysis can provide insights into the impact of longitudinal spin variations on vacancy formation at high temperatures. That is, due to the difference of spin magnitudes in the PM state between the atoms near the vacancy and in the bulk, a steady variation of $G_f^{mag}$ with temperature is expected, which is at variance with the predictions from classical Heisenberg models. In this case, it is also non-trivial to apply the well-known Ruch model [93], which proposes

$$\frac{G_f^{mag}(T) - G_f^{mag}(0K)}{G_f^{mag}(PM) - G_f^{mag}(0K)} = 1 - S^2 \quad (11)$$

where $S$ is the magnetic long-range order parameter, and $G_f^{mag}(PM)$ is the vacancy formation magnetic free energy in the ideal PM state. $G_f^{mag}(PM)$ can be taken as the saturated value for bcc Fe, but the definition of $G_f^{mag}(PM)$ is ambiguous for fcc Fe and Ni in which $G_f^{mag}(T)$ does not saturate at high temperatures.

### 4. Impact of magnetic transitions on vacancy formation

It has been shown in Fig. 4 that the effect of the magnetic transition is much stronger in bcc Fe than in fcc Fe and Ni. This is mainly related to the change in the exchange interaction energy across the magnetic transitions. To illustrate this, we begin with the CH models where $J_{ij}$ are the same for the atoms in the bulk and near the vacancy, and we focus on the vacancy formation energy $E_f^{mag}$ for simplicity. In this case, the exchange interaction energy $J_{ij}\boldsymbol{M}_i\boldsymbol{M}_j$ does not contribute to $E_f^{mag}$ in the ideal fully PM state. In the magnetic ground state, this contribution to $E_f^{mag}$ is equal to half the ground-state exchange interaction energy per atom. Therefore, the variation in $E_f^{mag}$ across the magnetic transition is equal to half the ground-state exchange interaction energy per atom. The latter value is expected to be larger in bcc Fe than in fcc Fe and Ni in view of the DFT results [84, 94], which show that the energy difference between various magnetic states is more significant in bcc Fe than in fcc Fe and Ni. For a quantitative comparison, we show in Table V the different contributions to the vacancy formation energies of the three systems. Although there is a compensating variation between the on-site (longitudinal) and exchange (transversal) contributions in the three systems, the latter contribution is much stronger in bcc Fe than in fcc Fe and Ni.

TABLE V. Contributions of different terms in the Hamiltonians to the vacancy formation energy in the magnetic ground state and the PM state. The PM value is obtained at 1500 K where the transversal contribution is negligible in all the three systems.

| Contribution | bcc Fe GS | bcc Fe PM | fcc Fe GS | fcc Fe PM | fcc Ni GS | fcc Ni PM |
|---|---|---|---|---|---|---|
| NM $\epsilon_i$ | 2.00 | 2.00 | 2.37 | 2.37 | 1.43 | 1.43 |
| On-site $A_i M_i^2 + B_i M_i^4$ | -0.07 | 0.00 | -0.47 | -0.53 | 0.14 | 0.06 |
| Exchange $J_{ij}\boldsymbol{M}_i\boldsymbol{M}_j$ | 0.27 | $\approx 0$ | -0.07 | $\approx 0$ | -0.13 | $\approx 0$ |
| Total | 2.20 | 2.00 | 1.84 | 1.85 | 1.43 | 1.48 |

## V. CONCLUSION

We investigated magnetic effects on vacancy formation properties in fcc Ni, and in bcc and fcc Fe. We performed density functional theory calculations and Monte Carlo simulations with effective interaction models.

DFT calculations were performed for a large set of magnetic configurations of the three systems, with or without a vacancy. These results reveal a dispersion of the vacancy formation energies $E_f$ among various magnetic states, which is larger in bcc and fcc Fe than in fcc Ni. The dispersion is related not only to the ordering of spin orientations but also to the spin magnitudes. We have shown that the spin ordering has a dominant influence on the vacancy formation energy in bcc Fe, while this formation energy in fcc Fe is more dependent on the spin magnitudes; in fcc Ni both types of effects are small.

The DFT-parametrized models, with a generalized-Heisenberg form, were used in Monte Carlo simulations enabling the thermal fluctuations of spin orientations and magnitudes. Regarding magnetic properties of the defect-free systems, the Monte Carlo results reveal that the longitudinal spin fluctuations are more significant in fcc Fe and Ni than in bcc Fe. In particular for fcc Ni, neglecting the temperature evolution of spin magnitudes leads to an overestimation by 220 K of the Curie point.

Based on an efficient Monte Carlo scheme, the vacancy formation energy, magnetic entropy and magnetic free energy were determined as functions of temperature for the three systems.

Overall, the predicted vacancy formation energies and equilibrium vacancy concentrations are in good agreement with the experimental data, which are available only at high temperatures. Concerning the determination of the equilibrium vacancy concentration, the vibrational entropy has a stronger contribution than the magnetic one for all the three systems. But the latter is nonnegligible in bcc Fe, as neglecting this contribution leads to an underestimation of the equilibrium vacancy concentration by a factor of 5.

The overall impact of the magnetic transition on the vacancy formation properties are found to be more significant in bcc Fe than in fcc Fe and Ni. The substantial decrease of the vacancy formation energy from the ferromagnetic to the paramagnetic regime in bcc Fe is mainly due to the transversal rather than the longitudinal spin excitations. This is coherent with the strong dependence of the formation energy on the spin-orientation ordering in bcc Fe as predicted by DFT. Consistently, a weaker dependence of the vacancy formation energy on the spin ordering in fcc Fe and Ni leads to a smaller variation of the vacancy properties below and above the magnetic transition.

We note a significant effect of longitudinal spin excitations on the magnetic free energy of vacancy formation in fcc Fe, resulting in its steady decrease above the Néel point. Also, below its Néel point, such effect is comparable but opposite to that of the transversal excitations. These are consistent with the DFT results in fcc Fe, which demonstrate a stronger dependence of vacancy formation energy on the spin magnitude rather than the spin ordering. Interestingly, it is noted that the predicted vacancy formation energy in the paramagnetic state is close to the AFD-state value, but it is 0.52 eV lower than the formation energy obtained with nonmagnetic fcc Fe. Although the latter has been used in some recent studies on diffusion properties in fcc Fe, our results suggest that the paramagnetic fcc Fe is better represented by the AFD state than the nonmagnetic state.

Regarding the vacancy formation in fcc Ni, the transversal spin excitations just below the Curie temperature lead to a sudden increase of the vacancy formation magnetic free energy, while the longitudinal spin fluctuations above the Curie temperature lead to a steady increase of this quantity.

The cases of fcc Fe and Ni reveal a relevant effect of longitudinal spin excitations on the vacancy formation properties, which can not be well captured by a classical Heisenberg model. Also, at variance with the bcc-Fe case, the widely used Ruch model cannot be applied in these systems to predict the temperature evolution of energetics of vacancy formation.

## ACKNOWLEDGMENTS

This work was performed using DARI-GENCI resources under the A0090906020 project, and the CINECA-MARCONI supercomputer within the SIS-TEEL project. The research leading to these results has been partially carried out in the framework of EERA Joint Program for Nuclear Materials and is partly funded by the European Commission HORIZON 2020 Framework Programme under Grant Agreement No. 755269. K. Li is supported by the CEA NUMERICS program, which has received funding from the European Union's Horizon 2020 research and innovation program under the Marie Skłodowska-Curie grant agreement No. 800945.